\documentclass[aps,prb,superscriptaddress,amssymb,amsmath,12pt,fullpage,tightenlines]{revtex4}
\usepackage{graphicx}
\usepackage{epstopdf}\usepackage{dcolumn}
\usepackage{bm}
\usepackage{epstopdf}
\usepackage{comment}
\usepackage{subfigure}
\usepackage{algpseudocode}
\usepackage{algorithm}
\usepackage[percent]{overpic}

\usepackage{enumitem}
\setlistdepth{9}
\setlist[enumerate,1]{label=$\bullet$}
\setlist[enumerate,2]{label=$\bullet$}
\setlist[enumerate,3]{label=$\bullet$}
\setlist[enumerate,4]{label=$\bullet$}
\setlist[enumerate,5]{label=$\bullet$}
\setlist[enumerate,6]{label=$\bullet$}
\setlist[enumerate,7]{label=$\bullet$}
\setlist[enumerate,8]{label=$\bullet$}
\setlist[enumerate,9]{label=$\bullet$}
\renewlist{enumerate}{enumerate}{9}

\begin{document}

\title{GPU Accelerated Discrete Element Method (DEM) Molecular Dynamics for Conservative, Faceted Particle Simulations}
\author{Matthew Spellings}
\affiliation{Chemical Engineering, University of Michigan}
\affiliation{Biointerfaces Institute, University of Michigan}
\author{Ryan L. Marson}
\affiliation{Materials Science \& Engineering, University of Michigan}
\affiliation{Biointerfaces Institute, University of Michigan}
\author{Joshua A. Anderson}
\affiliation{Chemical Engineering, University of Michigan}
\affiliation{Biointerfaces Institute, University of Michigan}
\author{Sharon C. Glotzer}
\affiliation{Chemical Engineering, University of Michigan}
\affiliation{Materials Science \& Engineering, University of Michigan}
\affiliation{Biointerfaces Institute, University of Michigan}
\date{\today}                                           

\begin{abstract}
Faceted shapes, such as polyhedra, are commonly found in systems of nanoscale, colloidal, and granular particles.
Many interesting physical phenomena, like crystal nucleation and growth, vacancy motion, and glassy dynamics are challenging to model in these systems because they require detailed dynamical information at the individual particle level.
Within the granular materials community the Discrete Element Method has been used extensively to model systems of anisotropic particles under gravity, with friction.
We provide an implementation of this method intended for simulation of hard, faceted nanoparticles, with a conservative Weeks-Chandler-Andersen (WCA) interparticle potential, coupled to a thermodynamic ensemble.
This method is a natural extension of classical molecular dynamics and enables rigorous thermodynamic calculations for faceted particles.

\end{abstract}

 \maketitle
\pagebreak
\pagenumbering{arabic}

\section{Introduction}

The impact of particle shape on the self-assembly of systems of colloidal- and nanoscale particles is receiving ever-increasing attention.\cite{Henzie2012}
Hard particle simulations are the most straightforward way to determine the impact of particle shape on assembly and have been highly successful in elucidating the phase behavior of anisotropic particles.\cite{Damasceno2012,Young2013}
Monte Carlo (MC) methods are ideal for probing the equilibrium behavior of such systems and can be implemented efficiently on modern highly parallel architectures.\cite{AndersonMPMC,Anderson2016hpmc}
However, nonequilibrium and/or dynamical properties often require deterministic, rather than stochastic, simulation methods.
Event-driven molecular dynamics\cite{Bannerman2011,Marfn1997,Miller2004,Smallenburg2012a} is one such method, but it can be difficult to parallelize or to extend for arbitrary shapes and can slow down at the moderate to high densities of interest in many self-assembly studies.

The Discrete Element Method (DEM) has been used extensively by the granular materials community to study dynamics of anisotropic, frictional particle systems.\cite{Ghaboussi1990,galindo2009molecular,Alonso-Marroquin2009a,Wang2010b,Mack2011a,Langston2013,boton2013}
This method models interactions between particles as interactions between the minimal set of lower-dimensional geometric features needed to capture the effects of particles' shapes.
DEM is also a natural method to implement as a force field in a classical molecular dynamics (MD) framework, which lends itself to hardware acceleration.
Programs using graphics processing units (GPUs) can achieve order-of-magnitude speedups over single-CPU programs, but only if they are made to take full advantage of the parallel nature of the GPU.
Here we present an adaptation of DEM to run on GPUs within the HOOMD-Blue\cite{Anderson2008} MD framework.


\section{model}




\begin{figure}
\begin{center}
\includegraphics[width=3in]{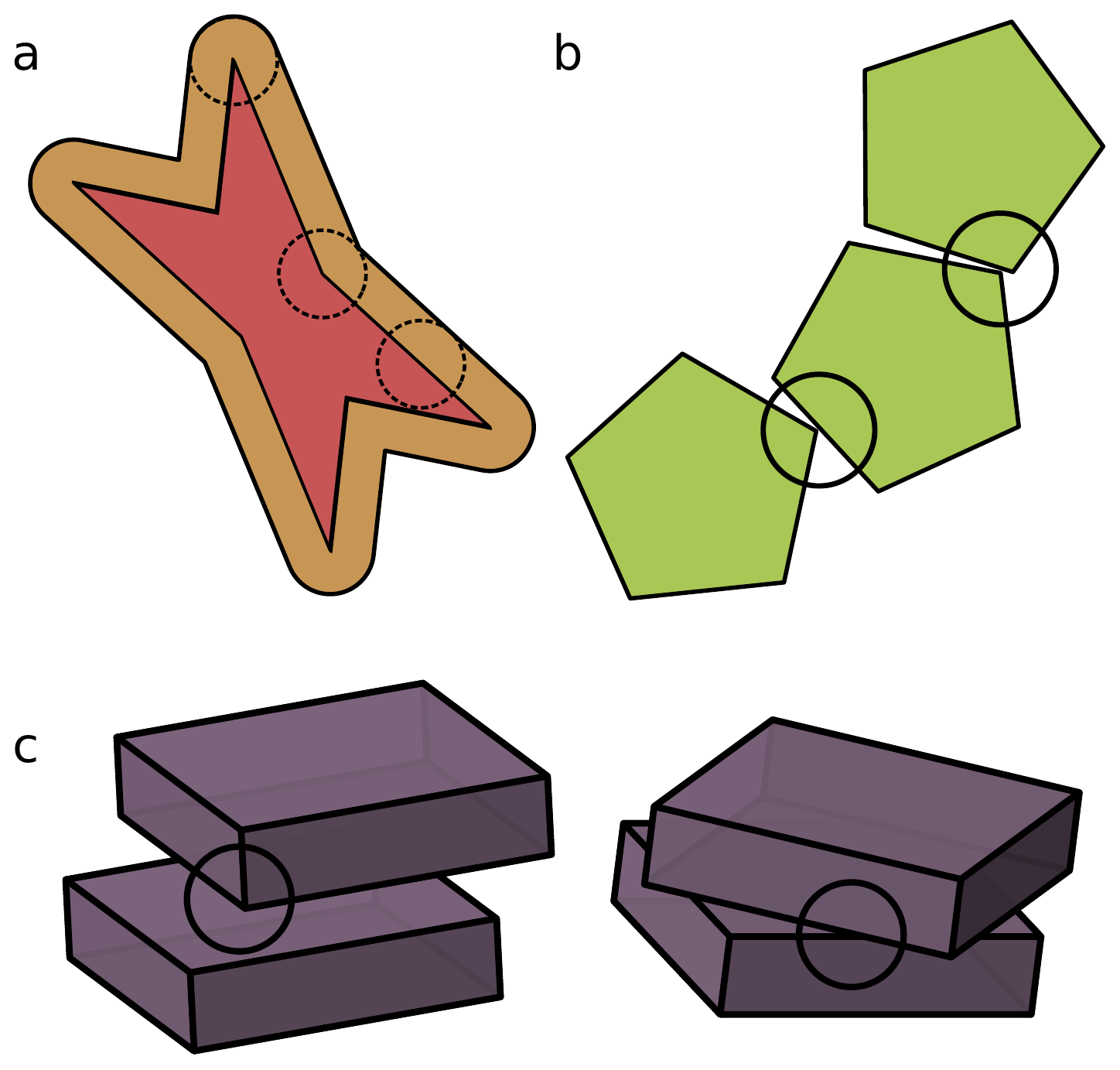}
\caption{The DEM model.
(a) Shapes are represented by a polygon or polyhedron (inner shape), rounded out by a disk or sphere (dotted circles) to give the outer shape. The inner shape can be concave or convex.
(b) In two dimensions, a repulsive contact force is evaluated between the nearest point on all pairs of vertices and edges.
(c) In three dimensions, the contact force is evaluated between all pairs of vertices and faces and all pairs of edges and edges.
}
\label{fig:Checks}
\end{center}
\end{figure}

Some assumptions must be made about the simulated particles for the method described here to be useful in classical molecular dynamics.
First, we assume that the polytopal shapes have been rounded by a sphere or disk of a given radius, as shown in Figure~\ref{fig:Checks}(a).
Rounding prevents discontinuities in the force arising from corners of shapes.
For studies of nanoparticle and colloidal assembly, we later show that the small amount of rounding required by our DEM algorithm has no discernible effect on self-assembly thermodynamics for rounding radii of up to 15\% of the edge length in some shapes.
Second, we assume that particles interact only through short-range, repulsive forces.
These assumptions allow us to approximate interactions between the volumes of particles $i$ and $j$ as interactions between the nearest points of lower-dimensional features (faces, edges, or vertices) of the particles using simple point-point interactions, without integrating over the surface or volume of the particles.
To prevent particle overlap in 2D, it is sufficient to consider interactions between all pairs of vertices and edges between the two particles, as shown in Figure~\ref{fig:Checks}(b).
In 3D, simulating polyhedral volumes requires calculating interactions between vertex-face pairs, and all edge-edge pairs as in Figure~\ref{fig:Checks}(c).

Often in MD and MC simulations of anisotropic particles, a single ``particle'' is built from several spheres, joined together rigidly\cite{escobedo2004,nguyen2011}; interactions are calculated between all pairs of spheres, but translations and rotations are applied to the body as a whole\cite{rahman1971}.
Ideally, to create an anisotropic potential from a shape, we would integrate an isotropic potential over the surface or volume of a pair of shapes.
Within DEM, particles are instead built up out of the geometric features of a 2D or 3D polytope: vertices ($V$), edges ($E$), and faces ($F$).
The functions $V(\cdot)$, $F(\cdot)$, and $E(\cdot)$ yield the coordinates of the vertices, faces, and edges of their argument, respectively, and $r_{ab}^*$ returns the distance between the nearest points of two features with types $a$ and $b$.
We then define the potential energy $U_{ij}$ between particles $i$ and $j$ in 3D using a point particle potential $\mathcal{U}$ as

\begin{equation}
\label{DEM}
U_{ij}^{3D} = \displaystyle\sum\limits_{\substack{E_i \in E(i) \\ E_j \in E(j)}} \mathcal{U}(r_{EE}^*( E_i,E_j)) +
    \displaystyle\sum\limits_{\substack{V_i \in V(i) \\ F_j \in F(j)}} \mathcal{U}(r_{VF}^*(V_i,F_j)) +
    \displaystyle\sum\limits_{\substack{V_j \in V(j) \\ F_i \in F(i)}} \mathcal{U}(r_{VF}^*(V_j,F_i))
\end{equation}

\vspace{8pt}

\noindent In 2D these features are reduced to checks between vertices and edges only:

\begin{equation}
\label{DEM_2D}
U_{ij}^{2D} = \displaystyle\sum\limits_{\substack{V_i \in V(i) \\ E_j \in E(j)}} \mathcal{U}(r_{VE}^*(V_i, E_j)) +
    \displaystyle\sum\limits_{\substack{V_j \in V(j) \\ E_i \in E(i)}} \mathcal{U}(r_{VE}^*(V_j,E_i))
\end{equation}

\vspace{8pt}

The nearest points given by $r_{ab}^*$ can be found using standard point-line, line-line, and point-plane formulae.
Forces are computed using the derivative of this potential and torques are based on the interaction point on each particle.

Because we are simulating nanoscale and colloidal systems, we choose a conservative pair potential $\mathcal{U}$ that is representative of the interactions of such materials and well vetted within the community.
A truncated and shifted version of the Lennard-Jones (LJ) potential, the Weeks-Chandler-Andersen (WCA) potential,\cite{Chandler1983} creates a steep, purely repulsive force from the particle surface with a rounding radius of $\frac{1}{2}\sigma_{ij}$:

\begin{equation}
\label{LJ}
U^{LJ}_{\mathrm{ij}}(r) = 4\epsilon_{\mathrm{ij}}\left[\left(\frac{\sigma_{\mathrm{ij}}}{r}\right)^{12}
-\left(\frac{\sigma_{\mathrm{ij}}}{r}\right)^{6}\right]
\end{equation}

\begin{equation}
U^{WCA}_{\mathrm{ij}}(r) = \left\{\begin{aligned}
	& U^{LJ}_{\mathrm{ij}}(r) - U^{LJ}_{\mathrm{ij}}(r^{WCA}_{\mathrm{cut}})
	& r  & < r^{WCA}_{\mathrm{cut}} \\
	& 0 & r  & \geqslant r^{WCA}_{\mathrm{cut}}
\end{aligned}\right.
\label{eqn:WCA}
\end{equation}

where $r^{WCA}_{\mathrm{cut}} = 2^{\frac{1}{6}}\sigma_{ij}$.

When initializing particles on a lattice, e.g. prior to thermalization, one may encounter collisions of perfectly parallel edges when the lattice spacing is small.
These collisions introduce a numerical instability for the molecular dynamics integrator: the points of interaction fluctuate at every timestep between the endpoints of each edge, yielding an unstable torque that changes sign at every timestep.
To alleviate this issue while still only using point interactions, when two edges are sufficiently close to parallel the interaction point is taken to be the midpoint of overlap between the two edges.

Another common occurrence during simulation is for features to be ``overcounted.''
This could lead to energetic ``bumps'' in the interaction: while the cutoff radius is not affected, the interaction is increased by some multiplicative factor according to the geometry of the two interacting sites, causing equipotential lines to expand slightly around vertices and edges.
This effect should not matter to the extent that the potential used is a good approximation of a ``hard'' force field.
As a concrete example, if two cubes are touching perfectly face to face, they will have an interaction strength 64 times as large as a single vertex-face interaction: 8 vertices are interacting with 3 faces each and 8 edges are interacting with 5 edges each.
If the single vertex-vertex interaction had a strength of $1 k_B T$, then the $1 k_B T$ isosurface for the now $64 k_B T$ interaction would have moved out by 10\% of the rounding radius of the shape.
Regardless of the geometry, we note that due to the cutoff in the WCA potential, it is impossible for the rounding radius to be increased by more than a factor of $2^{^1/_6} \approx 1.12$ with this overcounting effect.


\section{Algorithm}

\begin{figure}
\begin{center}
  \begin{overpic}[width=0.65\textwidth]{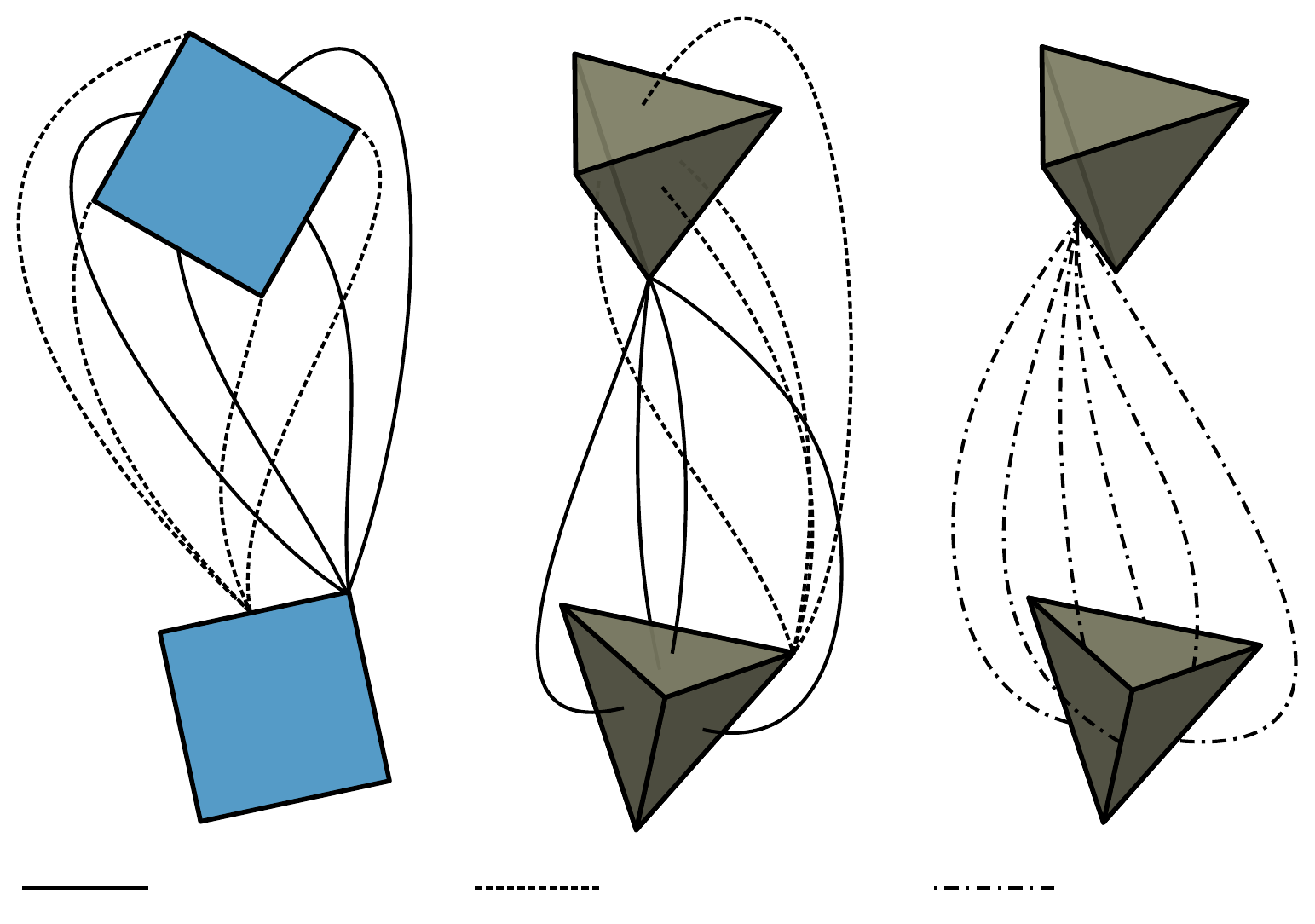}
    \put (12,2) {Thread 0}
    \put (46,2) {Thread 4}
    \put (81,2) {Thread 8}
\end{overpic}
\end{center}
\caption{Thread decomposition for the GPU. In 2D (left), each vertex is assigned a thread and each edge is assigned a thread; in 3D (center, right), each vertex is assigned two threads and each edge is assigned one thread.}
\label{fig:threadDecomposition}
\end{figure}

The total force, torque, and potential energy for a given particle is the sum of the force, torque, and potential energy contributions between it and its neighbors.
We evaluate these contributions from each particle's features independently by splitting the features among different GPU threads, then summing them efficiently in shared memory.
In 2D, two CUDA threads are assigned to each vertex of particle i, as shown in Figure~\ref{fig:threadDecomposition}.
The first thread assigned to a given vertex calculates and sums the force, torque, and potential energy contributions between that vertex and the nearest point to that vertex on each edge of each neighboring particle j.
The second thread assigned to a given vertex calculates and sums the force, torque, and potential energy contributions between the nearest point on the edge beginning at that vertex (travelling counterclockwise) in particle i to each vertex in each neighboring particle j.

In 3D, two CUDA threads are assigned to each vertex of particle i and one thread is assigned to each edge of particle i, as shown in Figure~\ref{fig:threadDecomposition}.
The first vertex thread calculates the interaction between that vertex and the nearest point to each face in each neighboring particle j.
The second vertex thread calculates the interaction between that vertex in each neighboring particle j and the nearest point of each face of particle i.
The edge thread calculates the interaction between the nearest point on its edge of particle i to each edge of each neighboring particle j.

\section{Results}

\subsection{Energy Conservation}

To perform rigorous thermodynamic calculations using this method, we must first ensure that the NVE integrator conserves energy when combined with our force algorithm.
We analyze both the short-term and long-term energy conservation of our model.\cite{allentildesley}
We use reduced units $\tau$ for time, $\epsilon$ for energy, and $m$ for mass.
We run NVE simulations of a fluid of squares (with edge length $4.24\sigma$) in 2D and tetrahedra (with edge length $8.49\sigma$) in 3D which have been thermalized at temperature $T^*=1$ in reduced units.

\begin{figure}
\begin{center}
\includegraphics[width=0.7\textwidth]{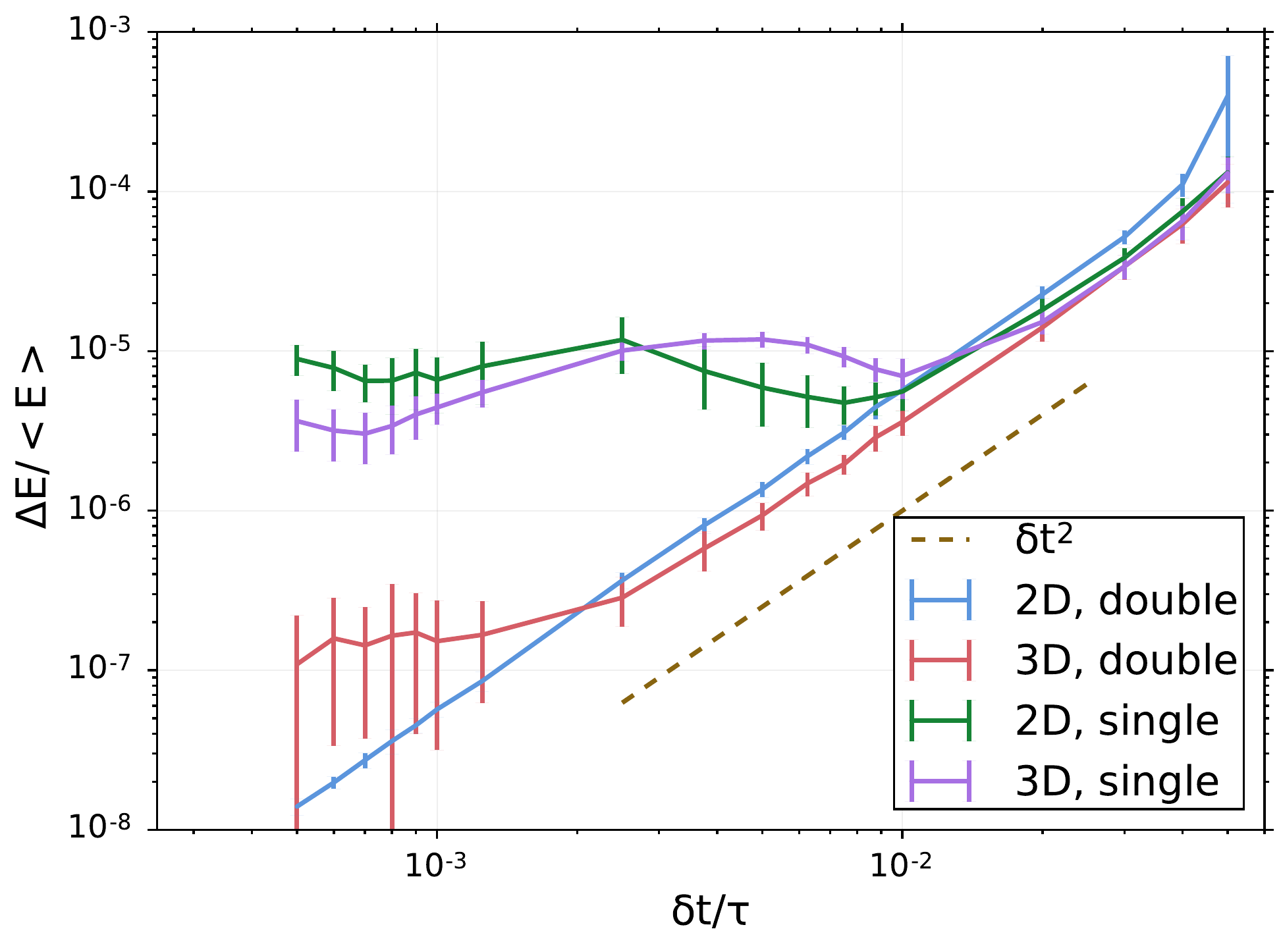}
\caption{Magnitude of total energy fluctuations for squares (2D) and tetrahedra (3D) in constant energy simulations for short times as a function of timestep size $\delta t$.}
\label{fig:energyFluctuations}
\end{center}
\end{figure}

To measure the short-term energy conservation, we compute the standard deviation of the total energy per particle $\sigma(E/N)$, recorded at a high frequency over a short NVE simulation of duration $10\tau$.
We measure $\sigma$ for systems using both single and double precision floating point arithmetic with a varying integration timestep size $\delta t$ and present the results in Figure~\ref{fig:energyFluctuations}.

For small $\delta t$, rounding errors saturate the accuracy of the method, causing the energy conservation to plateau.
These rounding errors appear at a much smaller value of $\delta t$ when using double- rather than single-precision floating point arithmetic.
At large $\delta t$, error is introduced through the coarse time step as $\frac{\Delta E}{<E>}\sim \Delta t^2$ for the second-order integrator we use\cite{Kamberaj2005}, increasing the energy deviation.
For the remaining tests, which are performed in single precision only, we choose $\delta t = 0.01 \tau$ as the timestep size to balance energy conservation and simulation speed.

To study the long-term energy conservation, we calculate the drift of the total energy per particle, $\frac{1}{N}\left( E(t) - E(0) \right)$, over long simulations.
For squares, we achieve an energy drift of $1.4\cdot10^{-5}~\Delta E/(N/\epsilon/\tau)$ and for tetrahedra we obtain $6.3\cdot10^{-4}~\Delta E/(N/\epsilon/\tau)$.
The energy drifts are significantly higher than those reported for isotropic particles\cite{anderson2013}, likely due to the new rotational degrees of freedom and the approximation of expressing the energy between features as the potential evaluated between their closest points.
In practice, we find these energy drifts to be acceptable for the coarse-grained simulations at which this method is targeted.

\subsection{Performance}

\begin{figure}
\begin{center}
\includegraphics[width=0.8\textwidth]{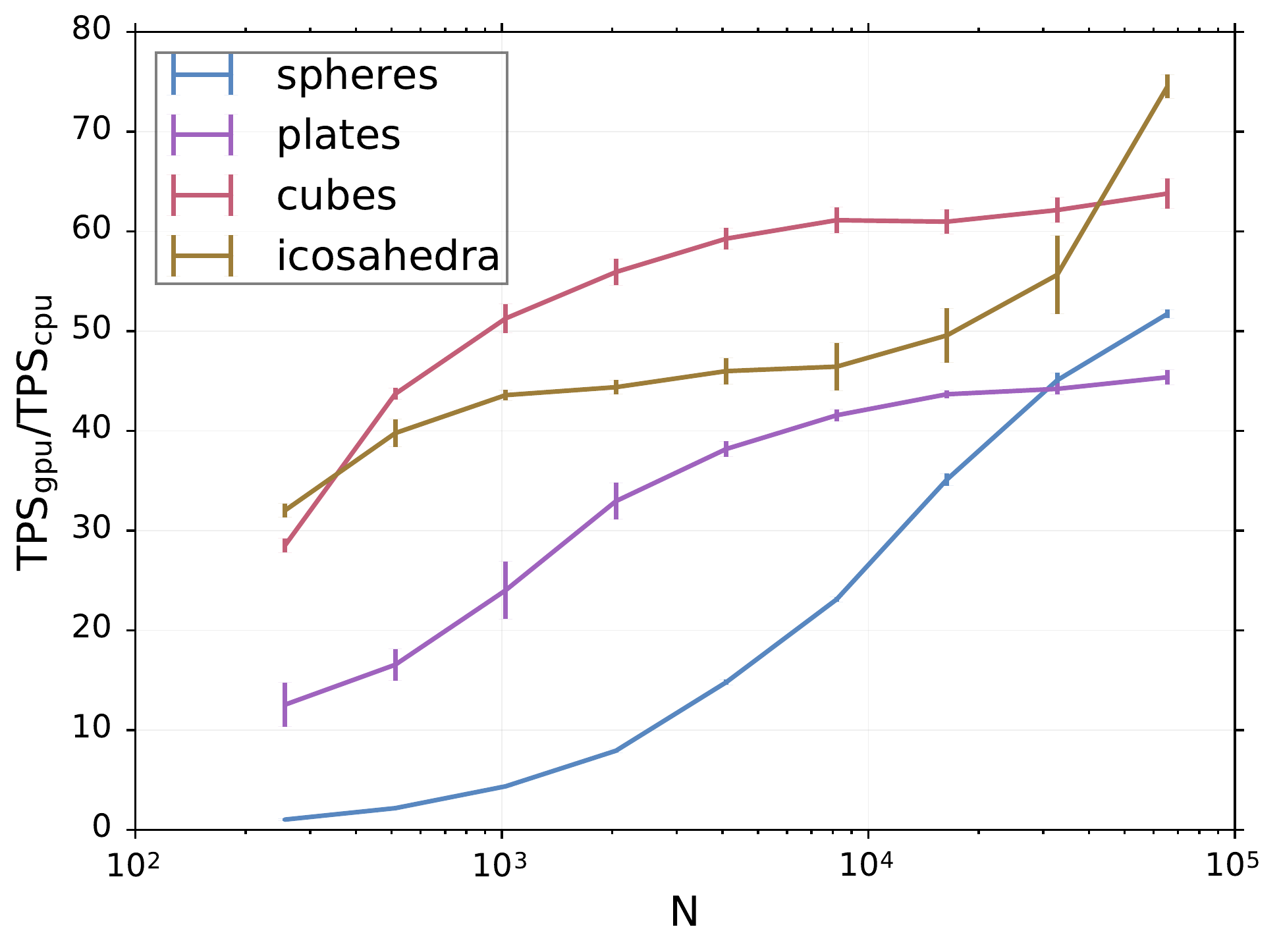}
\caption{Simulation speedup for different three-dimensional shapes by using a GPU relative to a single CPU core.}
\label{fig:performance}
\end{center}
\end{figure}

We evaluate the speed benefit of our GPU parallelization scheme using an NVIDIA Quadro M6000 relative to a single Intel E5-2680V2 CPU core for dense fluids of several systems with shapes of varying complexity in three dimensions: spheres, triangular plates, cubes, and icosahedra.
The anisotropic shapes are modeled with the DEM potential, and spheres are modeled with a central WCA interaction, as in Equation~\ref{eqn:WCA}.
As shown in Figure~\ref{fig:performance}, we achieve speedups of 15-75 times on the GPU, depending on the particle shape and system size.
For many shapes the relative speed saturates at system sizes of a few thousand particles; we note that the apparent jump in icosahedron performance is due to a decrease of speed on the CPU, likely due to memory locality effects, rather than an increase in GPU speed.
Typical absolute performance numbers, in particle-timesteps per second (PTPS), are reported in Table~\ref{tab:absolutePerformance}.
In contrast, spheres are only just beginning to saturate the GPU at 65,000 particles.
This finding demonstrates that the feature-based DEM parallelization scheme allows users to take advantage of GPU performance even for relatively small systems.

\begin{table}[h]
\begin{center}
\begin{tabular}{| l | c | c | c |} \hline
  Shape & $N$ & $PTPS_{CPU}$ & $PTPS_{GPU}$ \\ \hline
  Sphere & 256 & 3,200,000 & 3,000,000 \\ \hline
  Sphere & 4096 & 3,100,000 & 46,000,000 \\ \hline
  Sphere & 65536 & 3,100,000 & 159,000,000 \\ \hline
  Cube & 256 & 6,500 & 190,000 \\ \hline
  Cube & 4096 & 6,500 & 380,000 \\ \hline
  Cube & 65536 & 5,500 & 350,000 \\ \hline
  Icosahedron & 256 & 1,800 & 60,000 \\ \hline
  Icosahedron & 4096 & 1,800 & 80,000 \\ \hline
  Icosahedron & 65536 & 920 & 69,000 \\ \hline
\end{tabular}
\end{center}
\caption{Absolute performance of CPU and GPU implementations, in particle-timesteps per second (PTPS).}
\label{tab:absolutePerformance}
\end{table}

\subsection{Assembly}

\begin{figure}
\begin{center}
\includegraphics[width=\textwidth]{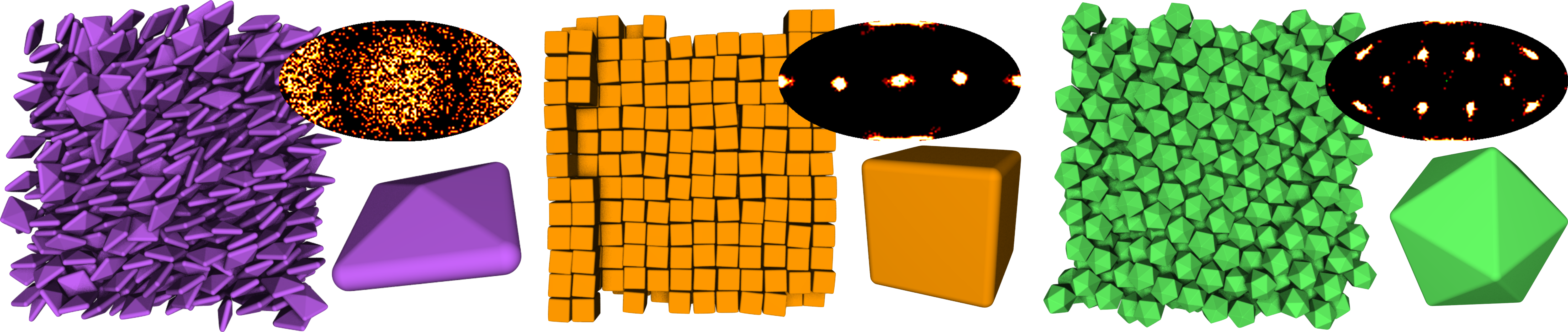}
\caption{Self-assembled crystal structures in systems of square bipyramids, cubes, and icosahedra.
Bond order diagrams depicting a global histogram of neighboring particle positions are in the upper-right of each snapshot.}
\label{fig:Assembly}
\end{center}
\end{figure}

We simulate the self-assembly of shapes into crystals to compare our results to those of hard particle colloidal crystals obtained \emph{via} MC simulations\cite{Damasceno2012}.
We simulate square bipyramids, cubes, and icosahedra in the NPT ensemble and over a range of pressures, as shown in Figure~\ref{fig:Assembly}.
We use a standard Nos\'e-Hoover thermostat applied to both translational and rotational degrees of freedom\cite{Kamberaj2005} and the barostat computes the pressure from the virial tensor, just as in MD\cite{Tuckerman2006}.
Similarly to MC results from hard particles with perfectly sharp corners, we find that the wide, flat bipyramids with a height of $\frac{1}{2 \sqrt{2}}$ relative to their edge length form a nematic phase (at $P=0.375\frac{k_BT}{\sigma^3}$), cubes form a simple cubic crystal (at $P=0.1\frac{k_BT}{\sigma^3}$), and icosahedra form a face-centered cubic crystal (at $P=0.05\frac{k_BT}{\sigma^3}$).

\section{Summary}

Here we have described a method to simulate conservative interactions of purely repulsive, rounded polytopes in molecular dynamics simulations.
Because evaluation of this potential is more intensive than that of a point particle, we are able to more efficiently utilize graphics processing units for smaller system sizes with this potential than with isotropic potentials.
Although the simulated shapes are still rounded, this method affords less opportunity for unphysical interdigitation than when rigid bodies of tangent or overlapping disks or spheres are used because the particle surface is smooth rather than rough.

In the future, there are many potentially useful optimizations that could be applied to this method.
Currently, the contact point search and pair potential evaluation happen within the same GPU kernel; in three dimensions in particular, this leads to large register usage and low GPU occupancy.
By splitting the contact search and force evaluation into two separate steps, some of this inefficiency could be avoided.
In general, the search for contact points could also be improved through the use of shape-local cell lists\cite{verlet1967}, octrees\cite{meagher1982}, or other spatial data structures.

In summary, the DEM-MD method presented here enables dynamical simulations of faceted particles.
With the particular integrators used, the method as implemented is most relevant to the simulation of nanoparticles and colloids in solvents where particle motion is dominated by inertia.
Our implementation will be available in the next major release, version 2.0, of HOOMD-Blue\footnote{HOOMD-blue web page: http://codeblue.umich.edu/hoomd-blue}, along with Brownian and Langevin thermostats for modeling non-inertial regimes.
Because the interactions are conservative, this method is useful for computation of thermodynamic quantities.
Having real dynamical information enables a more direct mapping to studies of nonequilibrium processes such as crystallization, vitrification, jamming, and self-assembly of active matter than Monte Carlo simulations.


\clearpage

\bibliography{references}

\begin{thebibliography}{27}
\expandafter\ifx\csname natexlab\endcsname\relax\def\natexlab#1{#1}\fi
\expandafter\ifx\csname bibnamefont\endcsname\relax
  \def\bibnamefont#1{#1}\fi
\expandafter\ifx\csname bibfnamefont\endcsname\relax
  \def\bibfnamefont#1{#1}\fi
\expandafter\ifx\csname citenamefont\endcsname\relax
  \def\citenamefont#1{#1}\fi
\expandafter\ifx\csname url\endcsname\relax
  \def\url#1{\texttt{#1}}\fi
\expandafter\ifx\csname urlprefix\endcsname\relax\def\urlprefix{URL }\fi
\providecommand{\bibinfo}[2]{#2}
\providecommand{\eprint}[2][]{\url{#2}}

\bibitem[{\citenamefont{Henzie et~al.}(2012)\citenamefont{Henzie, Gr{\"u}nwald,
  Widmer-Cooper, Geissler, and Yang}}]{Henzie2012}
\bibinfo{author}{\bibfnamefont{J.}~\bibnamefont{Henzie}},
  \bibinfo{author}{\bibfnamefont{M.}~\bibnamefont{Gr{\"u}nwald}},
  \bibinfo{author}{\bibfnamefont{A.}~\bibnamefont{Widmer-Cooper}},
  \bibinfo{author}{\bibfnamefont{P.~L.} \bibnamefont{Geissler}},
  \bibnamefont{and} \bibinfo{author}{\bibfnamefont{P.}~\bibnamefont{Yang}},
  \bibinfo{journal}{Nature Materials} \textbf{\bibinfo{volume}{11}},
  \bibinfo{pages}{131} (\bibinfo{year}{2012}), ISSN \bibinfo{issn}{1476-1122},
  \urlprefix\url{http://dx.doi.org/10.1038/nmat3178}.

\bibitem[{\citenamefont{Damasceno et~al.}(2012)\citenamefont{Damasceno, Engel,
  and Glotzer}}]{Damasceno2012}
\bibinfo{author}{\bibfnamefont{P.~F.} \bibnamefont{Damasceno}},
  \bibinfo{author}{\bibfnamefont{M.}~\bibnamefont{Engel}}, \bibnamefont{and}
  \bibinfo{author}{\bibfnamefont{S.~C.} \bibnamefont{Glotzer}},
  \bibinfo{journal}{Science} \textbf{\bibinfo{volume}{337}},
  \bibinfo{pages}{453} (\bibinfo{year}{2012}), ISSN \bibinfo{issn}{0036-8075},
  \urlprefix\url{http://www.sciencemag.org/cgi/doi/10.1126/science.1220869}.

\bibitem[{\citenamefont{Young et~al.}(2013)\citenamefont{Young, Personick,
  Engel, Damasceno, Barnaby, Bleher, Li, Glotzer, Lee, and Mirkin}}]{Young2013}
\bibinfo{author}{\bibfnamefont{K.~L.} \bibnamefont{Young}},
  \bibinfo{author}{\bibfnamefont{M.~L.} \bibnamefont{Personick}},
  \bibinfo{author}{\bibfnamefont{M.}~\bibnamefont{Engel}},
  \bibinfo{author}{\bibfnamefont{P.~F.} \bibnamefont{Damasceno}},
  \bibinfo{author}{\bibfnamefont{S.~N.} \bibnamefont{Barnaby}},
  \bibinfo{author}{\bibfnamefont{R.}~\bibnamefont{Bleher}},
  \bibinfo{author}{\bibfnamefont{T.}~\bibnamefont{Li}},
  \bibinfo{author}{\bibfnamefont{S.~C.} \bibnamefont{Glotzer}},
  \bibinfo{author}{\bibfnamefont{B.}~\bibnamefont{Lee}}, \bibnamefont{and}
  \bibinfo{author}{\bibfnamefont{C.~a.} \bibnamefont{Mirkin}},
  \bibinfo{journal}{Angewandte Chemie International Edition} pp.
  \bibinfo{pages}{n/a--n/a} (\bibinfo{year}{2013}), ISSN
  \bibinfo{issn}{14337851},
  \urlprefix\url{http://doi.wiley.com/10.1002/anie.201306009}.

\bibitem[{\citenamefont{Anderson et~al.}(2013)\citenamefont{Anderson,
  Jankowski, Grubb, Engel, and Glotzer}}]{AndersonMPMC}
\bibinfo{author}{\bibfnamefont{J.~A.} \bibnamefont{Anderson}},
  \bibinfo{author}{\bibfnamefont{E.}~\bibnamefont{Jankowski}},
  \bibinfo{author}{\bibfnamefont{T.~L.} \bibnamefont{Grubb}},
  \bibinfo{author}{\bibfnamefont{M.}~\bibnamefont{Engel}}, \bibnamefont{and}
  \bibinfo{author}{\bibfnamefont{S.~C.} \bibnamefont{Glotzer}},
  \bibinfo{journal}{Journal of Computational Physics}
  \textbf{\bibinfo{volume}{254}}, \bibinfo{pages}{27 } (\bibinfo{year}{2013}),
  ISSN \bibinfo{issn}{0021-9991},
  \urlprefix\url{http://www.sciencedirect.com/science/article/pii/S0021999113004968}.

\bibitem[{\citenamefont{Anderson et~al.}(2016)\citenamefont{Anderson, Irrgang,
  and Glotzer}}]{Anderson2016hpmc}
\bibinfo{author}{\bibfnamefont{J.~A.} \bibnamefont{Anderson}},
  \bibinfo{author}{\bibfnamefont{M.~E.} \bibnamefont{Irrgang}},
  \bibnamefont{and} \bibinfo{author}{\bibfnamefont{S.~C.}
  \bibnamefont{Glotzer}}, \bibinfo{journal}{Computer Physics Communications}
  (\bibinfo{year}{2016}), ISSN \bibinfo{issn}{0010-4655},
  \urlprefix\url{http://www.sciencedirect.com/science/article/pii/S001046551630039X}.

\bibitem[{\citenamefont{Bannerman et~al.}(2011)\citenamefont{Bannerman,
  Sargant, and Lue}}]{Bannerman2011}
\bibinfo{author}{\bibfnamefont{M.~N.} \bibnamefont{Bannerman}},
  \bibinfo{author}{\bibfnamefont{R.}~\bibnamefont{Sargant}}, \bibnamefont{and}
  \bibinfo{author}{\bibfnamefont{L.}~\bibnamefont{Lue}},
  \bibinfo{journal}{Journal of computational chemistry}
  \textbf{\bibinfo{volume}{32}}, \bibinfo{pages}{3329} (\bibinfo{year}{2011}),
  ISSN \bibinfo{issn}{1096-987X},
  \urlprefix\url{http://www.ncbi.nlm.nih.gov/pubmed/21953566}.

\bibitem[{\citenamefont{Marín}(1997)}]{Marfn1997}
\bibinfo{author}{\bibfnamefont{M.}~\bibnamefont{Marín}},
  \bibinfo{journal}{Computer Physics Communications}
  \textbf{\bibinfo{volume}{102}}, \bibinfo{pages}{81 } (\bibinfo{year}{1997}),
  ISSN \bibinfo{issn}{0010-4655},
  \urlprefix\url{http://www.sciencedirect.com/science/article/pii/S0010465597000118}.

\bibitem[{\citenamefont{Miller and Luding}(2004)}]{Miller2004}
\bibinfo{author}{\bibfnamefont{S.}~\bibnamefont{Miller}} \bibnamefont{and}
  \bibinfo{author}{\bibfnamefont{S.}~\bibnamefont{Luding}},
  \bibinfo{journal}{Journal of Computational Physics}
  \textbf{\bibinfo{volume}{193}}, \bibinfo{pages}{306} (\bibinfo{year}{2004}),
  ISSN \bibinfo{issn}{00219991},
  \urlprefix\url{http://linkinghub.elsevier.com/retrieve/pii/S0021999103004339}.

\bibitem[{\citenamefont{Smallenburg et~al.}(2012)\citenamefont{Smallenburg,
  Filion, Marechal, and Dijkstra}}]{Smallenburg2012a}
\bibinfo{author}{\bibfnamefont{F.}~\bibnamefont{Smallenburg}},
  \bibinfo{author}{\bibfnamefont{L.}~\bibnamefont{Filion}},
  \bibinfo{author}{\bibfnamefont{M.}~\bibnamefont{Marechal}}, \bibnamefont{and}
  \bibinfo{author}{\bibfnamefont{M.}~\bibnamefont{Dijkstra}},
  \bibinfo{journal}{Proceedings of the National Academy of Sciences of the
  United States of America} \textbf{\bibinfo{volume}{109}},
  \bibinfo{pages}{17886} (\bibinfo{year}{2012}), ISSN
  \bibinfo{issn}{1091-6490},
  \urlprefix\url{http://www.ncbi.nlm.nih.gov/pubmed/23012241}.

\bibitem[{\citenamefont{Ghaboussi and Barbosa}(1990)}]{Ghaboussi1990}
\bibinfo{author}{\bibfnamefont{J.}~\bibnamefont{Ghaboussi}} \bibnamefont{and}
  \bibinfo{author}{\bibfnamefont{R.}~\bibnamefont{Barbosa}},
  \bibinfo{journal}{International Journal for Numerical and Analytical Methods
  in Geomechanics} \textbf{\bibinfo{volume}{14}}, \bibinfo{pages}{451}
  (\bibinfo{year}{1990}).

\bibitem[{\citenamefont{Galindo-Torres
  et~al.}(2009)\citenamefont{Galindo-Torres, Alonso-Marroqu{\'\i}n, Wang,
  Pedroso, and Castano}}]{galindo2009molecular}
\bibinfo{author}{\bibfnamefont{S.}~\bibnamefont{Galindo-Torres}},
  \bibinfo{author}{\bibfnamefont{F.}~\bibnamefont{Alonso-Marroqu{\'\i}n}},
  \bibinfo{author}{\bibfnamefont{Y.}~\bibnamefont{Wang}},
  \bibinfo{author}{\bibfnamefont{D.}~\bibnamefont{Pedroso}}, \bibnamefont{and}
  \bibinfo{author}{\bibfnamefont{J.~M.} \bibnamefont{Castano}},
  \bibinfo{journal}{Physical Review E} \textbf{\bibinfo{volume}{79}},
  \bibinfo{pages}{060301} (\bibinfo{year}{2009}).

\bibitem[{\citenamefont{Alonso-Marroqu\'{\i}n and
  Wang}(2009)}]{Alonso-Marroquin2009a}
\bibinfo{author}{\bibfnamefont{F.}~\bibnamefont{Alonso-Marroqu\'{\i}n}}
  \bibnamefont{and} \bibinfo{author}{\bibfnamefont{Y.}~\bibnamefont{Wang}},
  \bibinfo{journal}{Granular Matter} \textbf{\bibinfo{volume}{11}},
  \bibinfo{pages}{317} (\bibinfo{year}{2009}), ISSN \bibinfo{issn}{1434-5021},
  \urlprefix\url{http://link.springer.com/10.1007/s10035-009-0139-1}.

\bibitem[{\citenamefont{Wang et~al.}(2010)\citenamefont{Wang, Yu, Langston, and
  Fraige}}]{Wang2010b}
\bibinfo{author}{\bibfnamefont{J.}~\bibnamefont{Wang}},
  \bibinfo{author}{\bibfnamefont{H.~S.} \bibnamefont{Yu}},
  \bibinfo{author}{\bibfnamefont{P.}~\bibnamefont{Langston}}, \bibnamefont{and}
  \bibinfo{author}{\bibfnamefont{F.}~\bibnamefont{Fraige}},
  \bibinfo{journal}{Granular Matter} \textbf{\bibinfo{volume}{13}},
  \bibinfo{pages}{1} (\bibinfo{year}{2010}), ISSN \bibinfo{issn}{1434-5021},
  \urlprefix\url{http://link.springer.com/10.1007/s10035-010-0217-4}.

\bibitem[{\citenamefont{Mack et~al.}(2011)\citenamefont{Mack, Langston, Webb,
  and York}}]{Mack2011a}
\bibinfo{author}{\bibfnamefont{S.}~\bibnamefont{Mack}},
  \bibinfo{author}{\bibfnamefont{P.}~\bibnamefont{Langston}},
  \bibinfo{author}{\bibfnamefont{C.}~\bibnamefont{Webb}}, \bibnamefont{and}
  \bibinfo{author}{\bibfnamefont{T.}~\bibnamefont{York}},
  \bibinfo{journal}{Powder Technology} \textbf{\bibinfo{volume}{214}},
  \bibinfo{pages}{431} (\bibinfo{year}{2011}), ISSN \bibinfo{issn}{00325910},
  \urlprefix\url{http://linkinghub.elsevier.com/retrieve/pii/S0032591011004529}.

\bibitem[{\citenamefont{Langston et~al.}(2013)\citenamefont{Langston, Ai, and
  Yu}}]{Langston2013}
\bibinfo{author}{\bibfnamefont{P.}~\bibnamefont{Langston}},
  \bibinfo{author}{\bibfnamefont{J.}~\bibnamefont{Ai}}, \bibnamefont{and}
  \bibinfo{author}{\bibfnamefont{H.-S.} \bibnamefont{Yu}},
  \bibinfo{journal}{Granular Matter} pp. \bibinfo{pages}{13--15}
  (\bibinfo{year}{2013}), ISSN \bibinfo{issn}{1434-5021},
  \urlprefix\url{http://link.springer.com/10.1007/s10035-013-0421-0}.

\bibitem[{\citenamefont{Boton et~al.}(2013)\citenamefont{Boton, Az\'ema,
  Estrada, Radja\"{\i}, and Lizcano}}]{boton2013}
\bibinfo{author}{\bibfnamefont{M.}~\bibnamefont{Boton}},
  \bibinfo{author}{\bibfnamefont{E.}~\bibnamefont{Az\'ema}},
  \bibinfo{author}{\bibfnamefont{N.}~\bibnamefont{Estrada}},
  \bibinfo{author}{\bibfnamefont{F.}~\bibnamefont{Radja\"{\i}}},
  \bibnamefont{and} \bibinfo{author}{\bibfnamefont{A.}~\bibnamefont{Lizcano}},
  \bibinfo{journal}{Phys. Rev. E} \textbf{\bibinfo{volume}{87}},
  \bibinfo{pages}{032206} (\bibinfo{year}{2013}),
  \urlprefix\url{http://link.aps.org/doi/10.1103/PhysRevE.87.032206}.

\bibitem[{\citenamefont{Anderson et~al.}(2008)\citenamefont{Anderson, Lorenz,
  and Travesset}}]{Anderson2008}
\bibinfo{author}{\bibfnamefont{J.}~\bibnamefont{Anderson}},
  \bibinfo{author}{\bibfnamefont{C.}~\bibnamefont{Lorenz}}, \bibnamefont{and}
  \bibinfo{author}{\bibfnamefont{A.}~\bibnamefont{Travesset}},
  \bibinfo{journal}{Journal of Computational Physics}
  \textbf{\bibinfo{volume}{227}}, \bibinfo{pages}{5342} (\bibinfo{year}{2008}),
  ISSN \bibinfo{issn}{00219991},
  \urlprefix\url{http://linkinghub.elsevier.com/retrieve/pii/S0021999108000818}.

\bibitem[{\citenamefont{John et~al.}(2004)\citenamefont{John, Stroock, and
  Escobedo}}]{escobedo2004}
\bibinfo{author}{\bibfnamefont{B.~S.} \bibnamefont{John}},
  \bibinfo{author}{\bibfnamefont{A.}~\bibnamefont{Stroock}}, \bibnamefont{and}
  \bibinfo{author}{\bibfnamefont{F.~A.} \bibnamefont{Escobedo}},
  \bibinfo{journal}{The Journal of Chemical Physics}
  \textbf{\bibinfo{volume}{120}}, \bibinfo{pages}{9383} (\bibinfo{year}{2004}),
  \urlprefix\url{http://scitation.aip.org/content/aip/journal/jcp/120/19/10.1063/1.1711594}.

\bibitem[{\citenamefont{Nguyen et~al.}(2011)\citenamefont{Nguyen, Jankowski,
  and Glotzer}}]{nguyen2011}
\bibinfo{author}{\bibfnamefont{T.~D.} \bibnamefont{Nguyen}},
  \bibinfo{author}{\bibfnamefont{E.}~\bibnamefont{Jankowski}},
  \bibnamefont{and} \bibinfo{author}{\bibfnamefont{S.~C.}
  \bibnamefont{Glotzer}}, \bibinfo{journal}{ACS Nano}
  \textbf{\bibinfo{volume}{5}}, \bibinfo{pages}{8892} (\bibinfo{year}{2011}),
  \bibinfo{note}{pMID: 21950837}, \eprint{http://dx.doi.org/10.1021/nn203067y},
  \urlprefix\url{http://dx.doi.org/10.1021/nn203067y}.

\bibitem[{\citenamefont{Rahman and Stillinger}(1971)}]{rahman1971}
\bibinfo{author}{\bibfnamefont{A.}~\bibnamefont{Rahman}} \bibnamefont{and}
  \bibinfo{author}{\bibfnamefont{F.~H.} \bibnamefont{Stillinger}},
  \bibinfo{journal}{The Journal of Chemical Physics}
  \textbf{\bibinfo{volume}{55}}, \bibinfo{pages}{3336} (\bibinfo{year}{1971}),
  \urlprefix\url{http://scitation.aip.org/content/aip/journal/jcp/55/7/10.1063/1.1676585}.

\bibitem[{\citenamefont{Chandler et~al.}(1983)\citenamefont{Chandler, Weeks,
  and Andersen}}]{Chandler1983}
\bibinfo{author}{\bibfnamefont{D.}~\bibnamefont{Chandler}},
  \bibinfo{author}{\bibfnamefont{J.}~\bibnamefont{Weeks}}, \bibnamefont{and}
  \bibinfo{author}{\bibfnamefont{H.}~\bibnamefont{Andersen}},
  \bibinfo{journal}{Science} \textbf{\bibinfo{volume}{220}},
  \bibinfo{pages}{787} (\bibinfo{year}{1983}),
  \urlprefix\url{http://www.sciencemag.org/content/220/4599/787.short}.

\bibitem[{\citenamefont{Allen and Tildesley}(1989)}]{allentildesley}
\bibinfo{author}{\bibfnamefont{M.~P.} \bibnamefont{Allen}} \bibnamefont{and}
  \bibinfo{author}{\bibfnamefont{D.~J.} \bibnamefont{Tildesley}},
  \emph{\bibinfo{title}{Computer simulation of liquids}}
  (\bibinfo{publisher}{Oxford university press}, \bibinfo{year}{1989}).

\bibitem[{\citenamefont{Kamberaj et~al.}(2005)\citenamefont{Kamberaj, Low, and
  Neal}}]{Kamberaj2005}
\bibinfo{author}{\bibfnamefont{H.}~\bibnamefont{Kamberaj}},
  \bibinfo{author}{\bibfnamefont{R.~J.} \bibnamefont{Low}}, \bibnamefont{and}
  \bibinfo{author}{\bibfnamefont{M.~P.} \bibnamefont{Neal}},
  \bibinfo{journal}{The Journal of chemical physics}
  \textbf{\bibinfo{volume}{122}}, \bibinfo{pages}{224114}
  (\bibinfo{year}{2005}).

\bibitem[{\citenamefont{Anderson and Glotzer}(2013)}]{anderson2013}
\bibinfo{author}{\bibfnamefont{J.~A.} \bibnamefont{Anderson}} \bibnamefont{and}
  \bibinfo{author}{\bibfnamefont{S.~C.} \bibnamefont{Glotzer}},
  \emph{\bibinfo{title}{The development and expansion of hoomd-blue through six
  years of gpu proliferation}} (\bibinfo{year}{2013}),
  \eprint{arXiv:1308.5587}.

\bibitem[{\citenamefont{Tuckerman et~al.}(2006)\citenamefont{Tuckerman,
  Alejandre, L\'{o}pez-Rend\'{o}n, Jochim, and Martyna}}]{Tuckerman2006}
\bibinfo{author}{\bibfnamefont{M.~E.} \bibnamefont{Tuckerman}},
  \bibinfo{author}{\bibfnamefont{J.}~\bibnamefont{Alejandre}},
  \bibinfo{author}{\bibfnamefont{R.}~\bibnamefont{L\'{o}pez-Rend\'{o}n}},
  \bibinfo{author}{\bibfnamefont{A.~L.} \bibnamefont{Jochim}},
  \bibnamefont{and} \bibinfo{author}{\bibfnamefont{G.~J.}
  \bibnamefont{Martyna}}, \bibinfo{journal}{Journal of Physics A: Mathematical
  and General} \textbf{\bibinfo{volume}{39}}, \bibinfo{pages}{5629}
  (\bibinfo{year}{2006}).

\bibitem[{\citenamefont{Verlet}(1967)}]{verlet1967}
\bibinfo{author}{\bibfnamefont{L.}~\bibnamefont{Verlet}},
  \bibinfo{journal}{Phys. Rev.} \textbf{\bibinfo{volume}{159}},
  \bibinfo{pages}{98} (\bibinfo{year}{1967}),
  \urlprefix\url{http://link.aps.org/doi/10.1103/PhysRev.159.98}.

\bibitem[{\citenamefont{Meagher}(1982)}]{meagher1982}
\bibinfo{author}{\bibfnamefont{D.}~\bibnamefont{Meagher}},
  \bibinfo{journal}{Computer Graphics and Image Processing}
  \textbf{\bibinfo{volume}{19}}, \bibinfo{pages}{129} (\bibinfo{year}{1982}).

\end{thebibliography}

\begin{acknowledgments}
  We thank Richmond Newman for helpful discussion of optimization techniques.
  This work was supported as part of the Center for Bio-Inspired Energy Science, an Energy Frontier Research Center funded by the US Department of Energy, Office of Science, Basic Energy Sciences under Award {DE}-{SC}0000989 and by the National Science Foundation, Division of Materials Research Award \# DMR 1409620.
  M.S. was also supported in part by a National Science Foundation, Integrative Graduate Education and Research Traineeship (IGERT) under Grant DGE 0903629.
  R.L.M. acknowledges support from the University of Michigan Rackham Merit Fellowship program.
  Computational resources and services were provided by Advanced Research Computing at the University of Michigan, Ann Arbor.
  The Glotzer Group at the University of Michigan is an NVIDA GPU Research Center.
  Hardware support by NVIDIA Corp. is gratefully acknowledged.
\end{acknowledgments}

\clearpage





\end{document}